\documentclass[12pt]{article}
\usepackage[english]{babel}
\usepackage[utf8]{inputenc}
\usepackage[T1]{fontenc}
\usepackage{listings}
\usepackage{amsbsy}
\usepackage{amsfonts}
\usepackage{amsmath}
\usepackage{amssymb}
\usepackage{amsthm}
\usepackage{graphicx}
\usepackage{subcaption}
\usepackage[pdftex]{color}
\usepackage{dsfont}
\usepackage{enumerate}
\usepackage{float}
\usepackage{geometry, calc, color, setspace}%
\usepackage{indentfirst}
\usepackage{longtable}
\usepackage{multirow}
\usepackage{natbib}
\usepackage{xr-hyper}
\usepackage{hyperref,bookmark}

\newtheorem{theorem}{Theorem}[section]
\newtheorem{definition}{Definition}[section]

\newtheorem{remark}[theorem]{Remark}
\newtheorem{assumption}[theorem]{Assumption}

\hypersetup{
	colorlinks=true,
	linkcolor=blue,
	citecolor=red,
	filecolor=blue,
	urlcolor=blue,
}

\protect

\begin{document}

	\def\spacingset#1{\renewcommand{\baselinestretch}%
		{#1}\small\normalsize} \spacingset{1}


	\title{\bf Semiparametric time series models driven by latent factor}
	\author{\normalsize{\bf Gisele O. Maia}$^\star$\footnote{E-mail: giseleemaia07@gmail.com},\, {\bf Wagner Barreto-Souza}$^\sharp$$^\star$\footnote{E-mail: wagner.barretosouza@kaust.edu.sa},\, {\bf Fernando S. Bastos}$^\flat$\footnote{E-mail: fernando.bastos@ufv.br}\, and\, {\bf Hernando Ombao$^\sharp$\footnote{E-mail: hernando.ombao@kaust.edu.sa}}\vspace{.3cm}\\  	
		{\small\it $^\star$Departamento de Estat\'istica, Universidade Federal de Minas Gerais, Belo Horizonte, Brazil}\\
		{\small\it $^\sharp$Statistics Program, King Abdullah University of Science and Technology, Thuwal, Saudi Arabia}\\
		\small\it $^\flat$Instituto de Ci\^encias Exatas e Tecnol\' ogicas, Universidade Federal de Vi\c cosa, Florestal, Brazil}
	\maketitle

	\begin{abstract}
		We introduce a class of semiparametric time series models by assuming a quasi-likelihood approach driven by a latent factor process. More specifically, given the latent process, we only specify the conditional mean and variance of the time series and enjoy a quasi-likelihood function for estimating parameters related to the mean. This proposed methodology has three remarkable features: (i) no parametric form is assumed for the conditional distribution of the time series given the latent process; (ii) able for modelling non-negative, count, bounded/binary and real-valued time series; (iii) dispersion parameter is not assumed to be known. Further, we obtain explicit expressions for the marginal moments and for the autocorrelation function of the time series process so that a method of moments can be employed for estimating the dispersion parameter and also parameters related to the latent process. Simulated results aiming to check the proposed estimation procedure are presented. Real data analysis on unemployment rate and precipitation time series illustrate the potencial for practice of our methodology.
	\end{abstract}
	
	\vfill
	
	\section{Introduction}
	
	\citet{cox1981} characterized two classes of models for time-dependent data, observation-driven and parameter-driven models. Let $\left\{Y_{t}\right\}_{t\in\mathbb N}$ denote a time series along this section. In observation-driven approach, it is assumed some conditional distribution for $Y_t$ given $\mathcal F_t\equiv \sigma\{Y_{t-1},Y_{t-2},\ldots\}$. Some works on this kind of process are \citet{zeger1988}, \citet{benjamin2003}, \citet{davis2003}, \citet{rochabeta} and \cite{davliu2016}, among others.
	
	The focus of this present paper is time series driven by latent factor, which is one example of parameter-driven model. Here, it is assumed that $\left\{Y_{t}\right\}_{t\in\mathbb N}$ given a latent process $\left\{\alpha_{t}\right\}_{t\in\mathbb N}$ is conditionally independent but not identically distributed; a regression structure is considered for modelling the mean of the process. 
	
	A pioneering work is due to \cite{zeger}, where a semiparametric count time series model was proposed. In that model, given the latent process, the author specifies only the two first moments of the conditional distribution of the counts. Estimation of the parameters related to the mean for this semiparametric model is performed through a quasi-likelihood function and a method of moments is considered for estimating parameters related to the latent process.
	
	\citet{davis2000} studied the Poisson count time series model driven by a Gaussian latent factor with focus on checking the existence of such a process in practical situations. They  proposed estimation of the regression coefficients based on a generalized linear model (GLM) approach and established conditions for consistency and asymptotic normality of the GLM estimators. An explicit form for the asymptotic covariance of these estimators was also obtained.

	Another important work is due to \citet{davis2009}, where an one-parameter exponential family is considered for the response time series given the latent process so extending the paper by \citet{davis2000}. The authors focused on the negative binomial case (with known dispersion parameter and so belonging to this family) and derived asymptotic properties of the GLM estimators. Estimation of the parameters related to the latent factor is done based on a kind of ordinary least squares method.
	
	In this paper we introduce a class of semiparametric time series models by assuming a quasi-likelihood approach driven by a latent factor process. To do this, we only specify the conditional mean and variance of the time series given the latent process so enjoying a quasi-likelihood function for estimating parameters related to the mean. This proposed methodology has three remarkable features: (i) no parametric form is assumed for the conditional distribution of the time series given the latent process; (ii) able for modelling non-negative, count, bounded/binary and real-valued time series; (iii) dispersion parameter is not assumed to be known. Our semiparametric class has as particular cases the models introduced and studied by \cite{zeger}, \cite{davis2000} and \cite{davis2009}.

	Quasi-likelihood approach has been used in Time Series Analysis, see for example \cite{zeger1988}, \cite{heyde1997}, \cite{berkes2003}, \cite{francq2004}, \cite{straumann2006}, \cite{christou2014} and \cite{christou2015}. The models proposed in these papers are parametric and belong to the observation-driven class.
	
	This paper is organized in the following manner. In Section \ref{sec_def} we introduce our class of semiparametric time series models and discuss its application in detail for dealing with non-negative continuous, count, bounded/binary and $\mathbb R$-valued time series. Further, we obtain marginal moments and the autocorrelation function of the proposed models, which will be used for estimating parameters. Section \ref{sec_est} is devoted to the quasi-likelihood estimation for the mean parameters combined with a kind of method of moments for estimating the dispersion and parameters related to the latent process. We also provide a bootstrap strategy to obtain the standard errors of the parameter estimates. Monte Carlo simulations are addressed in Section \ref{sec_sim} to check finite-sample behaviour of the proposed estimators. Real data analysis on unemployment rate and precipitation time series are presented in Section \ref{sec_app} to illustrate the potencial for practice of our methodology. Concluding remarks are addressed in Section \ref{concl_rem}.
	
	\section{Model definition}\label{sec_def}
	
	In this section we define our class of semiparametric time series models and obtain some basic properties which will be useful for estimating parameters. Roughly speaking, our methodology consists in assuming a quasi-likelihood approach for the time series given a latent process. In this way, we are only requiring the first two moments of the conditional distribution. This follows in the direction of the paper by \cite{zeger} but here we are being more general by assuming a broad family of link mean and variance functions in contrast with that paper where these functions are setted to be the identity function.
	
	\begin{definition}\label{defsts}
		Let $\left \{ Y_{t} \right \}_{t\in\mathbb N}$ be a time series and $\left\{\alpha_{t}\right\}_{t\in\mathbb N}$ a latent stationary strongly mixing process. Our proposed class of semiparametric time series (STS) models is defined by the following specifications:
		\begin{eqnarray*}
			\begin{aligned}
				g(\widetilde{\mu}_{t}) & = x_{nt}^\top \beta + \alpha_{t}, \\
				E(Y_{t}|\alpha_{t}) & = \widetilde{\mu}_{t}, \\
				Var(Y_{t}|\alpha_{t})&  = \phi V(\widetilde{\mu}_{t}), 
			\end{aligned} 
		\end{eqnarray*}
		where $\beta = (\beta_{1},...,\beta_{q})^\top$ is the vector of regression coefficients, $ x_{nt} $ is a observable covariate vector (which can be depend on the sample size) with dimension $ q \times 1 $, $g(\cdot)$ is an invertible link function, $V(\cdot)$ is a variance function and $\phi>0$ is a dispersion parameter.
	\end{definition}
	
	\begin{remark}
		A remarkable feature of the STS models is their ability to deal with different kind of time series data. Due to the flexibility of the mean link and variance functions $g$ and $V$, we are able for modelling counts, positive continuous, bounded, binary and $\mathbb R$-valued time series.
	\end{remark}
	
	\begin{remark}
		The assumption on the latent process to be stationary and strongly mixing is important to obtain consistency and asymptotic normality of the generalized linear models estimators as discussed with details by \cite{davis2009}. The latent processes considered in this present paper satisfy these properties.
	\end{remark} 
	
	\cite{davis2000} and \cite{davis2009} proposed estimation of the parameters related to the mean through a generalized linear model (GLM) approach by ignoring the latent process. Under some conditions, they proved that the GLM estimators are consistent and asymptotic normal distributed. It is worth to mention that the standard errors cannot be obtained from the information matrix due to the GLM approach. The authors derived the correct information matrix for those models and proposed as an alternative to perform Monte Carlo simulation to obtain standard errors of the estimates. We will use this Monte Carlo simulation strategy here in this paper with some adaptations since we do not have an explicit conditional distribution for the time series given the latent process.  
	
	As discussed by \cite{zeger}, \cite{davis2000} and \cite{davis2009}, the following assumption is required to obtain consistency for estimating the parameter vector $\beta$.
	
	\begin{assumption}\label{restriction} Let $ \left\{ Y_{t} \right\}_{t\in\mathbb N}$ be as in Definition \ref{defsts}. We assume that the latent process $\left\{\alpha_{t}\right\}_{t\in\mathbb N}$ is such that 
		\begin{eqnarray*}
			E(Y_{t}) = E\left(h(x_{nt}^\top \beta + \alpha_{t})\right) = h(x_{nt}^\top \beta)
		\end{eqnarray*}
		for all $t\in\mathbb N$, where $h(\cdot)$ is the inverse of the link function $g(\cdot)$. 
	\end{assumption}
	
	\begin{remark}
		For more theoretical details on the Assumption \ref{restriction}, we recommend the paper by \cite{zeger}; see discussion in Subsection 3.1 after Eq. (4) from that paper.
	\end{remark}
	
	We now present the latent processes we will consider along this paper. Following the papers by \cite{zeger}, \cite{davis2000} and \cite{davis2009}, we assume a latent Gaussian AR(1) model for the count, positive continuous and real-valued cases. More explicitly, we have that 
	\begin{eqnarray}\label{gaussian}
	\alpha_t=c+\rho\alpha_{t-1}+\eta_{t},\quad t\in\mathbb N,
	\end{eqnarray}
	where $\{\eta_{t}\}\stackrel{i.i.d}{\sim}N(0,\sigma^2)$, $|\rho|<1$ and $c\in\mathbb R$ is an intercept chosen according Assumption \ref{restriction}. In this case, the process is well-known to be stationary and strongly mixing with marginals $\alpha_{t} \sim N \left( \frac{c}{1-\rho},\sigma^2 \right)$, for all $t\in\mathbb N$.

	We now discuss another latent process which will be used for the bounded and binary cases. 
	This will be based on the first-order gamma autoregression (with mean 1) proposed by \cite{sim1990}. 
	We say that a sequence $\{Z_t\}_{t\in\mathbb N}$ follows a first-order gamma autoregression (denoted by GAR(1)) if satisfies
	\begin{eqnarray*}
		Z_t=\kappa\odot Z_{t-1}+\eta_t,\quad t\in\mathbb N,\quad Z_0\sim \mbox{G}(1/\sigma^2,1/\sigma^2),
	\end{eqnarray*} 
	where the operator $\odot$ is defined by $\kappa\odot Z_{t-1}\stackrel{d}{=}\sum_{i=1}^{N_{t-1}}W_i$, with $N_{t-1}|Z_{t-1}=z\sim\mbox{Poisson}(\alpha\rho z)$, $\{W_i\}_{i=1}^\infty\stackrel{iid}{\sim}\mbox{Exponential}(\kappa)$ and  $\{\eta_t\}_{t=1}^\infty\stackrel{iid}{\sim}\mbox{G}(\sigma^2,\kappa)$ are assumed to be independent and $\kappa=\dfrac{1}{\sigma^2(1-\rho)}$, for $\sigma^2>0$ and $\rho\in(0,1)$. Here, $G(\sigma^2,\kappa)$ denotes a gamma distribution with shape and scale parameters $\sigma^2$ and $\kappa$, respectively. 
	
	The GAR process depends on the parameters $\sigma^2$ and $\rho$. The parameter $\rho$ controls the dependence of this process since that $\mbox{corr}(Z_{t+k},Z_t)=\rho^k$ for $t,k\in\mathbb N$. The marginals of this model are gamma distributed with mean 1 and variance $\sigma^2$, therefore the model is stationary; see \cite{sim1990}. The strong mixing property of this process was established recently by \cite{baromb2019}. Therefore, we define our latent process $\{\alpha_t\}_{t\in\mathbb N}$ in the bounded/binary case by
	\begin{eqnarray}\label{gar}
	\alpha_t=Z_t+\log E\left(\exp(-Z_t)\right)=Z_t-\dfrac{1}{\sigma^2}\log(1+\sigma^2), \quad t\in\mathbb N.
	\label{fcgama}
	\end{eqnarray}
	
	The shifted gamma process $\{\alpha_t\}_{t\in\mathbb N}$ above is necessary to satisfy Assumption \ref{restriction}. This will be clear when we deal with the bounded/binary case in Subsection \ref{bound_case}. A similar approach was considered by \cite{davis2009} for dealing with binary data. In that paper, the authors assumed a kind of shifted exponential process. Based on our approach, the shifted term is very simple in contrast with the term of the exponential process considered in \cite{davis2009} (see Experiment 2, page 743).
	
	In what follows, we define our semiparametric time series models in each situation by assuming some forms for the link and variance funtions. Marginal moments and the autocorrelation function are also provided.

	\subsection{Non-negative time series}\label{non-neg_case}
	
	Let $\{Y_{t}\}_{t\in\mathbb N}$ be a time series with support $\mathcal S\subset \mathbb R^+$ (non-negative real numbers). This case includes for example count and positive continuous time series. We consider a logarithm link function and a polynomial variance function $V(\mu) = \mu^{p}$, with $\mu>0$ and $p>0$. In this way, we define the model for non-negative time series as
	\begin{equation*}
	\begin{aligned}
	\log\widetilde{\mu}_{t} & = x_{nt}^\top \beta + \alpha_{t}, \\
	E(Y_{t}|\alpha_{t}) & = \widetilde{\mu}_{t} =  \exp(x_{nt}^\top \beta + \alpha_{t}) = \exp(x_{nt}^\top \beta) \epsilon_{t},  \\
	Var(Y_{t}|\alpha_{t}) & = \phi V(\widetilde{\mu}_{t}) = \phi \widetilde{\mu}_{t}^p,
	\end{aligned} 
	\end{equation*}
	where $\{\alpha_t\}_{t\in\mathbb N}$ is Gaussian AR(1) process defined in (\ref{gaussian}) and  $\epsilon_{t} = \exp\{\alpha_{t}\}$, for $t\in\mathbb N$. In order to ensure Assumption \ref{restriction} is in force, we need to take $E(\epsilon_{t})=1$. We have that $E(\epsilon_{t}) = \exp\left\{E(\alpha_{t})+0.5\mbox{Var}(\alpha_{t})\right\}=1$ implies $c = - \sigma^2(1-\rho)/2$. Consequently, $\alpha_{t} \sim N \left(-\sigma^2/2,\sigma^2\right)$, for all $t\in\mathbb N$.
	In this case, the sequence $\{\epsilon_t\}_{t\in\mathbb N}$ is a strictly stationary log-normal autoregressive model with mean $1$ and variance equal to $\sigma_{\epsilon}^2=\exp(\sigma^2)-1$. Its autocovariance and autocorrelation functions are given respectively by $$\gamma_\epsilon(k)\equiv\mbox{cov}(\epsilon_{t+k},\epsilon_{t})=\exp(\gamma(k))-1$$
	and
	$$\rho_\epsilon(k)\equiv\mbox{corr}(\epsilon_{t+k},\epsilon_{t})=\dfrac{\exp(\rho(k))-1}{\exp(\sigma^2)-1},$$
	where $\gamma(k)=\sigma^2\phi^k$ and $\rho(k)=\phi^k$ are the autocovariance and autocorrelation functions at lag $k\in\mathbb N$ of the process $\{\alpha_t\}_{t\in\mathbb N}$. The usage of this log-normal process on the Poisson regression is discussed in \cite{davis2000}.
	Under the above specifications, we obtain that the marginal mean and variance of $Y_{t}$ are
	\begin{equation*}
	\mu_t\equiv E(Y_{t}) = E(E(Y_{t}|\alpha_{t})) 
	= \exp(x_{nt}^\top \beta) E(\epsilon_{t}) 
	= \exp(x_{nt}^\top \beta) 
	\end{equation*}
	and
	\begin{equation}\label{var_non-neg}
	\begin{aligned}
	\mbox{Var}(Y_{t})  = E \left(\mbox{Var}(Y_{t}|\alpha_{t}) \right) + \mbox{Var}\left( E(Y_{t}|\alpha_{t}) \right) 
	= \phi E\left(\tilde{\mu}_{t}^{p} \right) + \mu_t^2\mbox{Var}\left(e^{\alpha_{t}}\right) 
	= \phi \mu_{t}^{p} (\sigma_{\epsilon}^2 +1)^{\frac{p(p-1)}{2}} +  \mu_{t}^2 \sigma_{\epsilon}^2. 
	\end{aligned}
	\end{equation}
	
	For $k>0$, the autocovariance function of $\{Y_t\}_{t\in\mathbb N}$ is
	\begin{equation}\label{cov_non-neg}
	\begin{aligned}
	\mbox{Cov}(Y_{t+k},Y_{t})  = \mbox{Cov}\left(E(Y_{t+k}|\alpha_{t+k}), E(Y_{t}|\alpha_{t})\right) +0
	= \mbox{Cov}\left(e^{x_{n,t+k}^\top \beta} e^{\alpha_{t+k}},e^{x_{nt}^{T} \beta} e^{\alpha_{t}}\right) 
	=  \mu_{t+k} \mu_{t} \gamma_{\epsilon}(k) 
	\end{aligned}
	\end{equation}
	and the autocorrelation function is given by
	\begin{equation*}
	\begin{aligned}
	\mbox{Corr}(Y_{t+k},Y_{t}) & = \frac{\mbox{Cov}(Y_{t+k},Y_{t})}{\sqrt{\mbox{Var}(Y_{t+k})\mbox{Var}(Y_{t})}} \\
	& = \frac{\mu_{t+k} \mu_{t} \gamma_{\epsilon}(k)}{\sqrt{\left[\phi \mu_{t+k}^{p} (\sigma_{\epsilon}^2 +1)^{\frac{p(p-1)}{2}} +  \mu_{t+k}^2 \sigma_{\epsilon}^2\right]\left[\phi \mu_{t}^p (\sigma_{\epsilon}^2 +1)^{\frac{p(p-1)}{2}} +  \mu_{t}^2 \sigma_{\epsilon}^2\right]}} \\
	& =  \frac{ \rho_{\epsilon}(k)}{\sqrt{\left[ \phi \sigma_{\epsilon}^{-2} \mu_{t+k}^{p-2}(\sigma_{\epsilon}^2 +1)^{\frac{p(p-1)}{2}} +1 \right]\left[ \phi \sigma_{\epsilon}^{-2} \mu_{t}^{p-2}(\sigma_{\epsilon}^2 +1)^{\frac{p(p-1)}{2}} +1 \right]}}.
	\end{aligned}
	\end{equation*}
	
	The model by \cite{zeger} is a particular case of the class discussed in this subsection by taking $V(\mu)=\mu$ ($p=1$) and $\phi=1$.
	
	\subsection{$\mathbb R$-valued time series}\label{R_case}
	
	Here we assume that the support of the sequence $\{Y_{t}\}_{t\in\mathbb N}$ is $\mathbb R$. We set an identity link function and variance function $V(\widetilde{\mu}_{t}) = 1$, so mimicking the first two moments of a normal distribution. More specifically, we have assumed that
	\begin{equation*}
	\begin{aligned}
	\widetilde{\mu}_{t}  & = x_{nt}^\top \beta + \alpha_{t},\\
	E(Y_{t}|\alpha_{t}) & = \widetilde{\mu}_{t} = x_{nt}^\top \beta + \alpha_{t},\\
	Var(Y_{t}|\alpha_{t}) & = \phi V(\widetilde{\mu}_{t}) = \phi, 
	\end{aligned}
	\end{equation*}
	where $\{\alpha_t\}_{t\in\mathbb N}$ is the Gaussian AR(1) process given in (\ref{gaussian}) with $c=0$, being this last condition necessary to ensure Assumption \ref{restriction} is in force. In this case, $\alpha_t\sim N(0,\sigma^2)$ for $t\in\mathbb N$.
	
	By using basic properties of conditional expectation, we obtain that the first two marginal cumulants of $Y_t$ are given by
	\begin{equation*}
	\mu_t\equiv E(Y_{t}) = E(E(Y_{t}|\alpha_{t})) = E(x_{nt}^\top \beta + \alpha_{t}) = x_{nt}^\top \beta
	\end{equation*}
	and
	\begin{equation}\label{var_Rcase}
	\begin{aligned}
	\mbox{Var}(Y_{t}) & = E \left(\mbox{Var}(Y_{t}|\alpha_{t}) \right) + \mbox{Var}\left( E(Y_{t}|\alpha_{t}) \right)  = \phi +\mbox{Var}\left(\mu_t + \alpha_{t}  \right)  = \phi + \sigma^2. 
	\end{aligned}
	\end{equation}
	
	The autocovariance and autocorrelation functions for $k>0$ are given by
	\begin{equation}\label{cov_Rcase}
	\begin{aligned}
	\mbox{Cov}(Y_{t+k},Y_{t})  = \mbox{Cov}\left(E(Y_{t+k}|\alpha_{t+k}), E(Y_{t}|\alpha_{t})\right)+0
	= \mbox{Cov} \left(\mu_{t+k} + \alpha_{t+k}, \mu_t + \alpha_{t} \right) 
	=   \sigma^2\rho(k)=\sigma^2\rho^k
	\end{aligned}
	\end{equation}
	and
	\begin{equation*}
	\begin{aligned}
	\mbox{Corr}(Y_{t+k},Y_{t})  = \frac{\mbox{Cov}(Y_{t+k},Y_{t})}{\sqrt{\mbox{Var}(Y_{t+k})\mbox{Var}(Y_{t})}}
	= \frac{\rho(k) \sigma^2}{\sqrt{(\phi + \sigma^2)^2}} 
	= \frac{\rho(k)}{\phi/\sigma^2+1}=\frac{\rho^k}{\phi/\sigma^2+1}. 
	\end{aligned}
	\end{equation*}
	
	A Gaussian time series model driven by a latent AR(1) process was considered by \cite{davis2009}; see Example 3 on page 742. In that model, the variance of the time series given the latent process is assumed known (so belonging to the one-parameter exponential family). In our proposed model here, no assumption on the distribution of the time series is imposed and its variance is assumed to be an unknown parameter to be estimated.
	
	\subsection{Bounded/binary time series}\label{bound_case}
	
	Let $\{Y_{t}\}_{t\in\mathbb N}$ be a process having one of the following supports: $(0,1)$, $\left\{0,1\right\}$ or $\{0,1,\ldots,m\}$, with $m\in\mathbb Z^+$. Therefore, here it is allowed proportions/rates (bounded continuous), binary and binomial time series data. We assume the link function to be $g(z)=-\log z$ and the variance function equal to $V(z) = z(1-z)$, for $z\in(0,1)$.
	
	Consider $\{\alpha_t\}_{t\in\mathbb N}$ be the shifted gamma process given in (\ref{gar}). Our model here is defined by the following equations:
	\begin{equation*}
	\begin{aligned}
	-\log\widetilde{\mu}_{t}  & = x_{nt}^\top \beta + \alpha_{t},\\
	E(Y_{t}|\alpha_{t}) & = \widetilde{\mu}_{t} =\exp(-x_{nt}^\top \beta)\epsilon_{t},\\
	Var(Y_{t}|\alpha_{t}) & = \phi V(\widetilde{\mu}_{t}) = \phi\widetilde{\mu}_{t}(1-\widetilde{\mu}_{t}), 
	\end{aligned}
	\end{equation*}
	where $\epsilon_{t}=\exp(-\alpha_t)$ and $\phi=1$ and $\phi=m$ for the binary and binomial cases, respectively. For the bounded continuous case, we have that $0<\phi<1$. Here, the vector $\beta$ is such that $x_{nt}^\top \beta>0$, since that $\widetilde{\mu}_{t}\in(0,1)$ for all $t\in\mathbb N$. The shifted term in the gamma autoregressive process is now justified. This is necessary to ensure that Assumption \ref{restriction} works. We have that the marginal mean of $Y_t$ is
	\begin{eqnarray*}
		\mu_t\equiv E(Y_t)=E(E(Y_t|\alpha_t))=\exp(-x_{nt}^\top \beta)E(\epsilon_t)=\exp(-x_{nt}^\top \beta),
	\end{eqnarray*}
	since $E(\epsilon_t)=E(\exp(-\alpha_t))=E(\exp(-Z_t))/E(\exp(-Z_t))=1$. After some algebra, we obtain that the marginal variance of $Y_t$ is 
	\begin{eqnarray}\label{varbound}
	\mbox{Var}(Y_t)=\phi\mu_t+\mu_t^2\left\{(1-\phi)\left(\dfrac{(1+\sigma^2)^2}{1+2\sigma^2}\right)^{1/\sigma^2}-1\right\}.
	\end{eqnarray}
	
	The autocovariance of the process $\{Y_{t}\}_{t\in\mathbb N}$ is
	\begin{equation*}
	\begin{aligned}
	\mbox{Cov}(Y_{t+k},Y_{t})  = \mbox{Cov}\left(E(Y_{t+k}|\alpha_{t+k}), E(Y_{t}|\alpha_{t})\right)+0
	= \mu_{t+k}\mu_t(1+\sigma^2)^{2/\sigma^2}\mbox{Cov}\left(\exp(-Z_{t+k}),\exp(-Z_t)\right),
	\end{aligned}
	\end{equation*}
	for $k>0$. From Eq. (2.6) from \cite{sim1990}, we have an explicit expression for the joint Laplace function of $(Z_{t+k},Z_t)$. Using that expression, we obtain that
	\begin{equation}\label{covbound}
	\begin{aligned}
	\mbox{Cov}(Y_{t+k},Y_{t})=\mu_{t+k}\mu_t\left\{\left(\dfrac{(1+\sigma^2)^2}{1+2\sigma^2+(\sigma^2)^ 2(1-\rho^k)}\right)^{1/\sigma^2}-1\right\}.
	\end{aligned}
	\end{equation}
	
	An expression for the autocorrelation function is immediately obtained by using (\ref{varbound}) and (\ref{covbound}). The time series model for binary data proposed here is an alternative to the model discussed by \cite{davis2009} since we are using a different latent process. We again call attention that the shifted term considered here is simpler than the term of that paper, which involves a multiplication of infinite number of terms. Further, our proposed methodology enables us to deal with continuous bounded time series data.
	
	\section{Quasi-likelihood approach and method of moments}\label{sec_est}
	
	In this section we discuss estimation of the parameters by combining quasi-likelihood approach and method of moments. Let $Y_1,\ldots,Y_n$ be a random trajectory of a time series process as in Definition \ref{sec_def} and $\theta = (\beta,\phi,\sigma^2,\rho)^\top$ be the parameter vector. Estimation of the parameter vector $\beta$ will be done through the quasi-likelihood method proposed by \cite{wedderburn1974}. The log-quasi-likelihood function is given by
	\begin{equation*}
	\mathcal Q(\beta) = \sum_{j=1}^nQ(y_j;\mu_j), 
	\end{equation*}
	where $Q(y;\mu)=\displaystyle\int_{y}^{\mu} \frac{y-u}{V(u)}du$. Depending on the choices for the variance function, the quasi-likelihood models have corresponding cases in the generalized linear models. These cases will be discussed in the following subsections. The quasi-likelihood estimator for $\beta$ is given by 
	\begin{eqnarray*}\label{betahat}
		\widehat\beta=\mbox{argmax}_{\beta}\mathcal Q(\beta).
	\end{eqnarray*}
	
	The quasi-likelihood estimate of $\beta$ can be obtained by using the \texttt{R} package \texttt{glm}. To estimate the remaining (nuisance) parameters, we use the moments and autocovariance function obtained in the previous section and then propose a kind of method of moments estimators. This strategy has been used for instance by \cite{zeger}, \cite{davis2000}, \cite{davis2009} and \cite{christou2014}.
	
	We obtain the standard errors for the quasi-likelihood estimates of $\beta$ through a Monte Carlo simulation. We call attention that we are not assuming a specific distribution for the time series given by the latent process. On the other hand, this is not a problem since it is enough in each Monte Carlo replica to assume a specific parametric model having the same mean structure of our semiparametric model, which works even under an incorrect specification of the variance function as argued by \cite{zeger}. It is worth to note that we are also interested in obtaining the standard errors for the nuisance parameter estimates, and then a correct specification of the variance function is required for this purpose. This procedure will be illustrated in the applications to real time series in Section \ref{sec_app}.
	
	In the following subsections we discuss estimation of the parameters with more details for the non-negative, real-valued and bounded/binary time series models.
	
	\subsection{Non-negative time series}\label{est_non-neg_case}
	
	Consider the non-negative time series model discussed in Subsection \ref{non-neg_case}. Then, we have the variance function given by $V(\mu) = \mu^p$ for $\mu,p>0$ and the marginal mean of $Y_t$ given by $\mu_{t} = \exp(x_{nt}^\top \beta)$, for $t=1,\ldots,n$. For $p\neq1,2$, we have that
	\begin{equation*}
	\begin{aligned}
	Q(y;\mu) & =  \int_{y}^{\mu} \frac{y-u}{u^{p}} du \\
	& =   \frac{y}{1-p} (\mu^{-p+1} - y^{-p+1}) - \frac{1}{2-p} (\mu^{-p+2}-y^{-p+2}). 
	\end{aligned}
	\label{qs1}
	\end{equation*}
	
	For $p=1$ and $p=2$, we obtain respectively $Q(y;\mu)=y(\log\mu-\log y)+y-\mu$ and $Q(y;\mu)=\log(y/\mu)-y/\mu+1$. The quasi-likelihood models with $p=1$, $p=2$ and $p=3$ have the Poisson, gamma and inverse-Gaussian generalized linear models as corresponding cases.
	
	To estimate $\phi$, $\sigma^2$ and $\rho$ through method of moments, we use the expressions of $\mbox{Var}(Y_{t})$ and $\mbox{Cov}(Y_{t + k}, Y_{t})$ (for $k=1,2$) given respectively in (\ref{var_non-neg}) and (\ref{cov_non-neg}) so obtaining
	\begin{equation}\label{phi_non-neg}
	\widehat{\phi} = \frac{\sum_{t=1}^{n}(Y_{t}-\widehat{\mu}_{t})^2 - (e^{\widehat\sigma^2}-1) \sum_{t=1}^{n}\widehat{\mu}_{t}^2}{ e^{\widehat\sigma^2p(p-1)/2}\sum_{t=1}^{n}\widehat{\mu}_{t}^{p}} 
	\end{equation}
	and
	\begin{equation}\label{eqs_non-neg}
	\exp(\widehat\sigma^2\widehat\rho^k) =  \frac{\sum_{t=1}^{n-k}(Y_{t} - \widehat{\mu}_{t})(Y_{t+k} - \widehat{\mu}_{t+k})}{\sum_{t=1}^{n-k}\widehat\mu_{t}\widehat\mu_{t+k}}+1,\quad\mbox{for}\,\, k=1,2,
	\end{equation}
	where $\widehat\mu_{t} = \exp(x_{nt}^\top\widehat\beta)$ for $t=1,\ldots,n$ with $\widehat\beta$ being the quasi-likelihood estimator of $\beta$. 
	
	Define $\mathcal M_k\equiv\log\left(\sum_{t=1}^{n-k}(Y_{t} - \widehat{\mu}_{t})(Y_{t+k}-\widehat{\mu}_{t+k})/\sum_{t=1}^{n-k}\widehat\mu_{t}\widehat\mu_{t+k}+1\right)$, for $k=1,2$. After some algebra, we obtain an explicit solution from the Equations given in (\ref{eqs_non-neg}), that is $\widehat\rho=\mathcal M_2/\mathcal M_1$ and $\widehat\sigma^2=\mathcal M^2_2/\mathcal M_1$. Consequently, we also obtain an explicit estimator for $\phi$ given in (\ref{phi_non-neg}).
	
	\subsection{$\mathbb R$-valued time series}\label{est_R_case}
	
	For real-valued time series, we have assumed that $V(\mu) = 1$. As discussed in Subsection \ref{R_case}, we choose the latent factor having null mean so that the marginal mean of $Y_t$ is $\mu_{t}=x_{nt}^\top\beta$. The $Q$-function is this case is given by
	\begin{equation*}
	\begin{aligned}
	Q(y;\mu) & = \int_{y}^{\mu} (y-u)  du \\
	& =  y\mu - \frac{\mu^2}{2} - \frac{y^2}{2}, 
	\end{aligned}
	\end{equation*}
	for $y,\mu\in\mathbb R$. By maximizing the logarithm of the quasi-likelihood function, we obtain the estimators for the regression coefficients, say $\widehat\beta$. From expressions (\ref{var_Rcase}) and (\ref{cov_Rcase}), we obtain the following method of moments estimators for $\phi$, $\rho$ and $\sigma^2$:
	\begin{equation*}
	\begin{aligned}
	\widehat{\phi}=\frac{1}{n} \sum_{t=1}^n (Y_{t}-\widehat\mu_t)^2-\widehat\sigma^2,  
	\end{aligned}
	\end{equation*}
	\begin{equation*}
	\widehat{\rho}=\dfrac{\sum_{t=1}^{n-2} (Y_{t}-\widehat\mu_t)(Y_{t+2}-\widehat\mu_{t+2})}{\sum_{t=1}^{n-1} (Y_{t}-\widehat\mu_t)(Y_{t+1}-\widehat\mu_{t+1})}
	\end{equation*}
	and
	\begin{equation*}
	\widehat{\sigma}^2=\dfrac{\left(\sum_{t=1}^{n-1} (Y_{t}-\widehat\mu_t)(Y_{t+1}-\widehat\mu_{t+1})\right)^2}{n\sum_{t=1}^{n-2} (Y_{t}-\widehat\mu_t)(Y_{t+2}-\widehat\mu_{t+2})}.
	\end{equation*}

	\subsection{Bounded/binary time series}\label{est_bound_case}
	
	In this case, for $y\in(0,1)$, the $Q$-function assumes the form
	\begin{eqnarray*}
		Q(y;\mu)&=&\displaystyle\int_{y}^{\mu} \frac{y-u}{V(u)}du\\
		&=&y\left\{\log\left(\mu(1-\mu)\right)-\log\left(\dfrac{y}{1-y}\right)\right\}+\log(1-\mu)-\log(1-y),
	\end{eqnarray*}
	
	For $y=0$ and $y=1$ we obtain respectively $Q(0,\mu)=\log(1-\mu)$ and $Q(1,\mu)=\log\mu$.
	Let $\widehat\mu_t=\exp(-x_{nt}^\top\widehat\beta)$, for $t=1,\ldots,n$, with $\widehat\beta$ denoting the quasi-likelihood estimator obtained based on the above $Q$-function. Assume $\phi$ is a unknown parameter to be estimated. From Expressions (\ref{varbound}) and (\ref{covbound}), we obtain that the method of moments estimator of $\phi$ is
	\begin{eqnarray*}\label{estboundphi}
		\widehat\phi=\dfrac{\sum_{t=1}^n(Y_t-\widehat\mu_t)^2-\left(w(\widehat\sigma^2)-1\right)\sum_{t=1}^n\widehat\mu_t^2}{\sum_{t=1}^n\widehat\mu_t-w(\widehat\sigma^2)\sum_{t=1}^n\widehat\mu_t^2}
	\end{eqnarray*}
	and the estimators for $\sigma^2$ and $\rho$ are obtained by solving the system of non-linear equations
	\begin{eqnarray}\label{estboundlat}
	v(\widehat\sigma^2,\widehat\rho^k)=\dfrac{\sum_{t=1}^{n-k}(Y_t-\widehat\mu_t)(Y_{t+k}-\widehat\mu_{t+k})}{\sum_{t=1}^{n-k}\widehat\mu_t\widehat\mu_{t+k}}+1,\quad \mbox{for}\,\, k=1,2,
	\end{eqnarray}
	where $w(x)=\left(\dfrac{(1+x)^2}{1+2x}\right)^{1/x}$ and $v(x,y)=\left(\dfrac{(1+x)^2}{1+2x+x^2(1-y)}\right)^{1/x}$, for $x>0$ and $y\in(0,1)$. Since there is not closed form for the method of moments estimators of $\sigma^2$ and $\rho$, some numerical optimization is needed. For the case where $\phi$ is known, as in the Bernoulli and binomial cases where $\phi=1$ and $\phi=m$, respectively, just use (\ref{estboundlat}) to get estimators for $\sigma^2$ and $\rho$. 
	
	\section{Simulated results}\label{sec_sim}
	
	We perform three simulation studies to evaluate the methodology presented for estimating the model parameters based on the quasi-likelihood approach combined with method of moments. All the implementations in this paper were conducted through the \cite{sor} software. We here illustrate the positive continuous, real-valued and bounded cases. For all cases considered in these simulated studies, we take $1000$ Monte Carlo replicas and sample sizes $n=500,1000,2000$.
	
	For the first case, we consider the semiparametric time series (STS) model for positive continuous data defined in Subsection \ref{non-neg_case} driven by the Gaussian AR(1) process. More specifically, we take the variance function to be quadratic, $V(\mu)=\mu^2$, so mimicking the GLM gamma model. In this simulation, we set the covariate vector $$x_{nt} = \left\{1,\cos(2\pi t/12),\sin(2\pi t/12) \right\}, \quad t=1,\ldots,n,$$ with regression coefficients $\beta = (5,-0.2,0.4)^\top$, $\phi=0.1$, $\sigma^2 = 0.5$ and $\rho=0.6$. For generating the simulated time series in each Monte Carlo replica, we assume a conditional gamma distribution (given the latent process) with mean $\widetilde\mu_t$ and variance $\phi V(\widetilde\mu_t)$, for $t=1,\ldots,n$. Estimation of the parameters is performed as proposed in Subsection \ref{est_non-neg_case}.
	
	In Table \ref{posit_sim}, we present the empirical means and standard errors of the quasi-likelihood estimates of the $\beta$'s and the method of moments (MMs) estimates of $\phi$, $\sigma^2$ and $\rho$ with their respective standard errors. We call attention that MM estimators can produce estimates out of the parameter space. In these cases, the samples were discarted and a new Monte Carlo replica was considered. This is a well-known problem of this kind of estimator and it is attenuated when working with moderate or large sample sizes. 
	
	
	\begin{table}
		\centering
		\caption{Empirical means and standard errors of the quasi-likelihood estimates of $\beta$ and method of moments estimates of $\phi$, $\sigma^2$ and $\rho$ based on the STS for positive continuous data.}\label{posit_sim} 
		\begin{tabular}{cccccccccc}
			\hline
			&&\multicolumn{2}{c}{$n=500$}&&
			\multicolumn{2}{c}{$n=1000$}&&
			\multicolumn{2}{c}{$n=2000$}\\
			\cline{3-4}\cline{6-7}\cline{9-10}
			parameter&true value&mean&stand. err.&& mean& stand. err.&& mean &stand. err.\\
			\hline
			$\beta_0$ &    5 &4.997   &0.070 &&4.998   &0.049&&4.997   &0.035 \\
			$\beta_1$ &$-$0.2&$-$0.199&0.076 &&$-$0.202&0.054&&$-$0.200&0.037 \\
			$\beta_2$ &0.4   &0.394   &0.074 &&0.398   &0.053&&0.401   &0.039 \\
			$\phi$    &0.1   &0.131   &0.089 &&0.115   &0.071&&0.107   &0.059 \\
			$\sigma^2$&0.5   &0.448   & 0.107&&0.475   &0.086&&0.487   & 0.058\\
			$\rho$    & 0.6  &0.626   &0.101 &&0.615   &0.075&&0.603   &0.102 \\
			\hline
		\end{tabular}
	\end{table}

\begin{figure}\centering
		\includegraphics[scale=0.3]{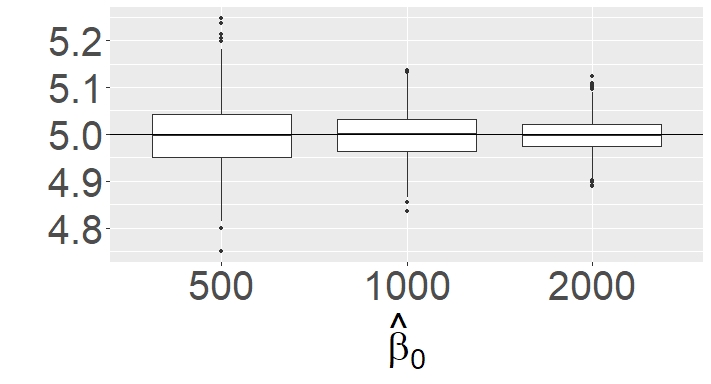}~\includegraphics[scale=0.3]{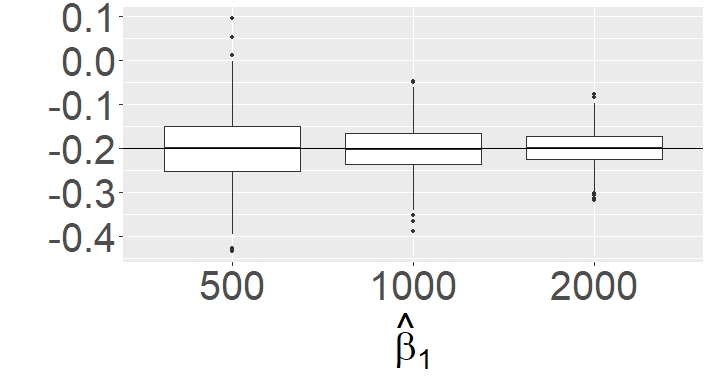}~\includegraphics[scale=0.3]{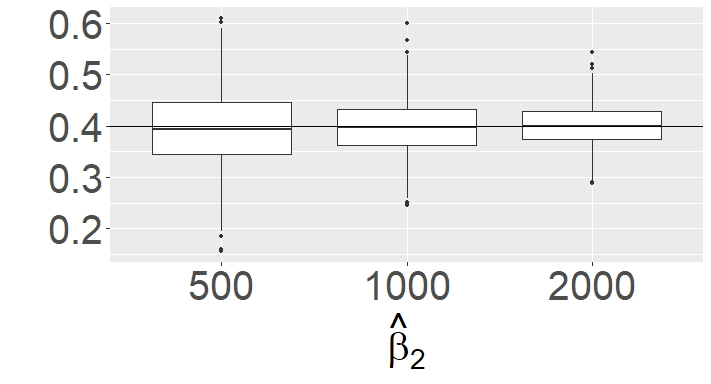}
		\includegraphics[scale=0.3]{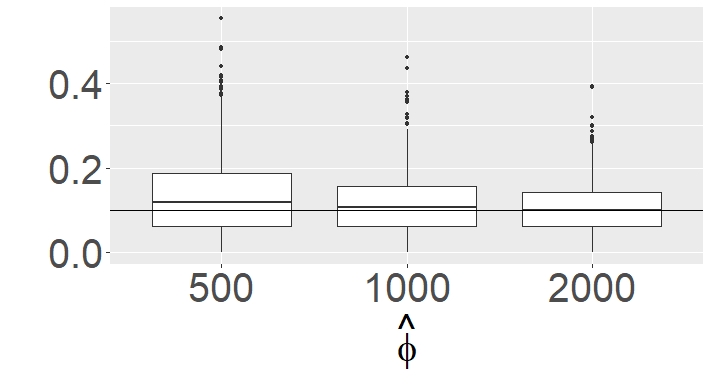}~\includegraphics[scale=0.3]{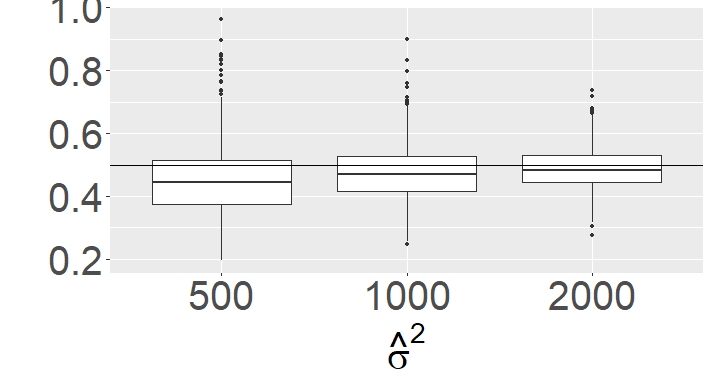}~\includegraphics[scale=0.3]{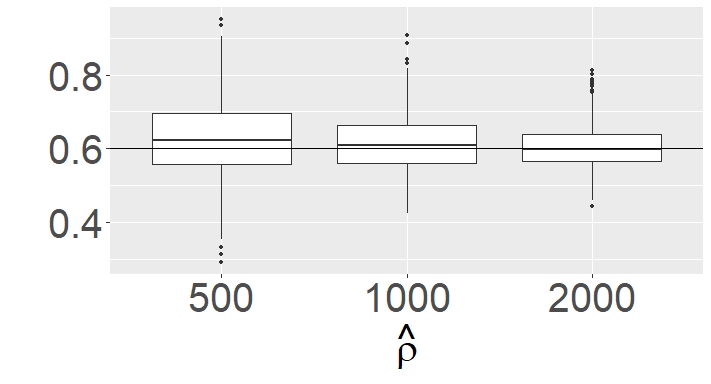}
		\caption{Boxplots of the parameter estimates based on the STS for positive continuous data.}
		\label{boxplot_posit}
	\end{figure}

	From Table \ref{posit_sim}, we observe that the quasi-likelihood estimators yielded almost unbiased estimates for $\beta$ for all sample sizes considered. The MM estimators also provided satisfactory results for estimating $\phi$, $\sigma^2$ and $\rho$. These comments are also supported from Figure \ref{boxplot_posit}, where boxplots of the parameter estimates are displayed. From these plots, we observe a general good performance and consistency of the proposed estimators as the sample size increases.
	
	We now consider a second scenario involving real-valued time series with the semiparametric model given in Subsection \ref{R_case} ($V(\mu) = 1$) based on the null mean Gaussian AR(1) process.
	
	We assume the covariate vector $$x_{nt} = \left\{  1, t/n,\cos(2\pi t/6) \right\}$$ with $\beta = (0.1,0.5,0.7)^\top$. We also set $\phi=3$, $\sigma^2 = 1$ and $\rho = 0.5$. To generate the $\mathbb R$-valued time series, we take the conditional distribution of $Y_t$ given the latent process to be normal distributed, for $t=1,\ldots,n$. 
	
	
	\begin{table}
		\centering
		\caption{Empirical means and standard errors of the quasi-likelihood estimates of $\beta$ and method of moments estimates of $\phi$, $\sigma^2$ and $\rho$ based on the STS for real-valued data.}\label{R_sim} 
		\begin{tabular}{cccccccccc}
			\hline
			&&\multicolumn{2}{c}{$n=500$}&&
			\multicolumn{2}{c}{$n=1000$}&&
			\multicolumn{2}{c}{$n=2000$}\\
			\cline{3-4}\cline{6-7}\cline{9-10}
			parameter&true value&mean&stand. err.&& mean& stand. err.&& mean &stand. err.\\
			\hline
			$\beta_0$ &0.1   &0.106 &0.218 &&0.100 &0.152 && 0.096&0.109 \\
			$\beta_1$ &0.5   & 0.496&0.382 &&0.501 &0.267 &&0.502 & 0.192\\
			$\beta_2$ & 0.7  &0.696 &0.126 &&0.697 &0.086 &&0.699 &0.060 \\
			$\phi$    & 3    &2.700 &0.810 &&2.813 &0.686 &&2.832 &0.560 \\
			$\sigma^2$& 1    &1.280 &0.800 &&1.184 &0.685 &&1.157 &0.555 \\
			$\rho$    & 0.5  &0.519 &0.230 &&0.516 &0.203 &&0.499 &0.174 \\
			\hline
		\end{tabular}
	\end{table}

\begin{figure}\centering
		\includegraphics[scale=0.3]{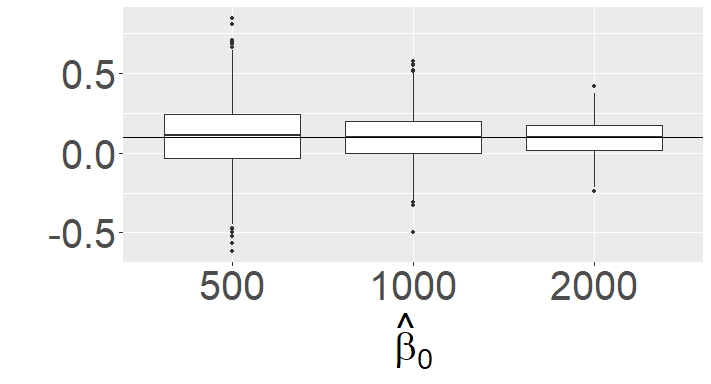}~\includegraphics[scale=0.3]{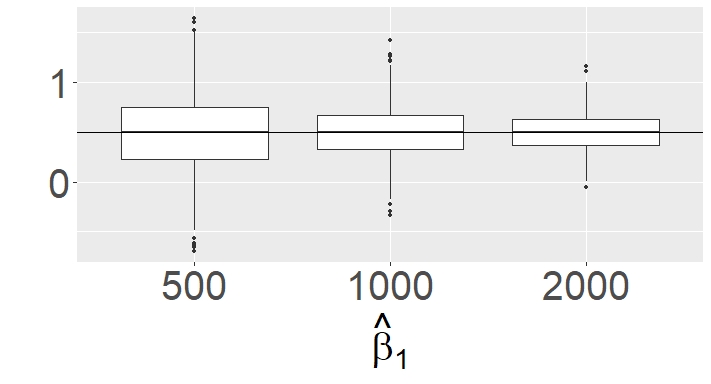}~\includegraphics[scale=0.3]{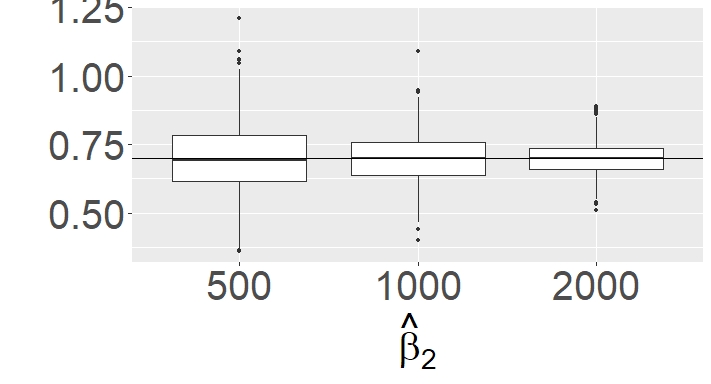}
		\includegraphics[scale=0.3]{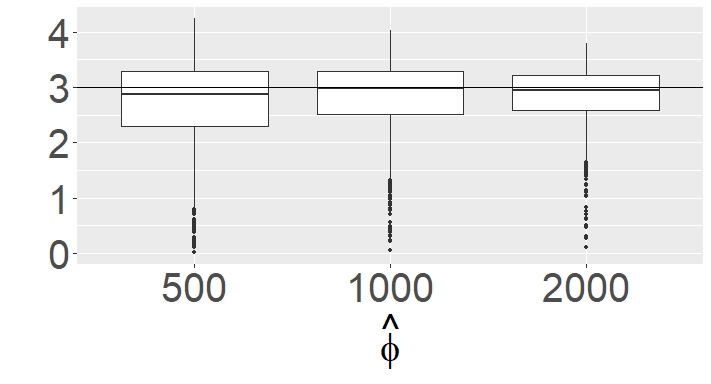}~\includegraphics[scale=0.3]{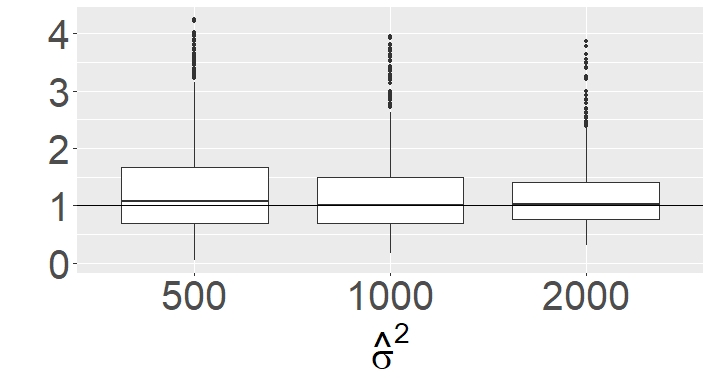}~\includegraphics[scale=0.3]{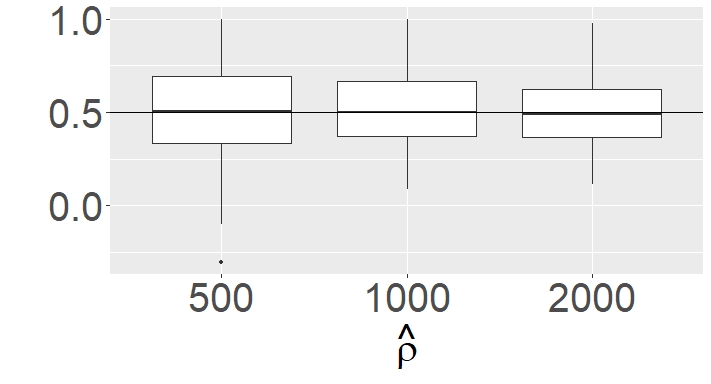}
		\caption{Boxplots of the parameter estimates based on the STS for real-valued data.}
		\label{boxplot_R}
	\end{figure}

	The empirical means and standard errors of the model parameters based on the estimation procedure discussed in Subsection \ref{est_R_case} are presented in Table \ref{R_sim}. Boxplots of these estimates obtained via Monte Carlo simulation are given in Figure \ref{boxplot_R}. From these results, we can observe a good performance of the proposed estimators based on quasi-likelihood approach combined with method of moments for the considered real-valued time series.
	
	Our last scenario is about bounded time series on the interval $(0,1)$. We illustrate the finite-sample behaviour of the estimators given in Subsection \ref{est_bound_case} for the bounded/binary STS model (presented in Subsection \ref{bound_case}) driven by the shifted gamma AR(1) process (\ref{gar}).
	
	Here, we set the parameters of the latent process as $\sigma^2 =0.3$ and $\rho=0.8$, the dispersion parameter $\phi=0.1$, and the regression covariates including trend components:
	$$x_{nt} = \left\{1, t/n, (t/n)^2 \right\}^\top,\quad t=1,\ldots,n,$$
	with associated regression coefficients $\beta = (1, 0.3, 0.5)^\top$. The time series generator here assumes that $Y_t$ given $\alpha_{t}$ follows a beta distribution with mean $\widetilde{\mu}_{t} = \exp(-x_{nt}^\top\beta-\alpha_{t})$ and variance $\phi\widetilde\mu_t(1-\widetilde\mu_t)$, for $t=1,\ldots,n$.
	
	In Table \ref{R_bound} we present the empirical means and standard errors of the quasi-likelihood estimates, as well as the estimates by the method of moments. In this case, we can observe a considerable bias in the quasi-likelihood estimates for the $\beta$'s, specially for $n=500$. On the other hand, we see a good performance of the method of moments estimators for the parameters $\phi$, $\sigma^2$ and $\rho$. This difficulty in estimating the regression coefficients was reported by \cite{davis2009} in a similar setting. The authors considered a binary time series model driven by an exponential latent process. In the simulated results of that paper, it is only assumed an intercept for the mean and a GLM approach is considered for estimating it, which yielded estimates with considerable bias.
	
\begin{figure}\centering
		\includegraphics[scale=0.3]{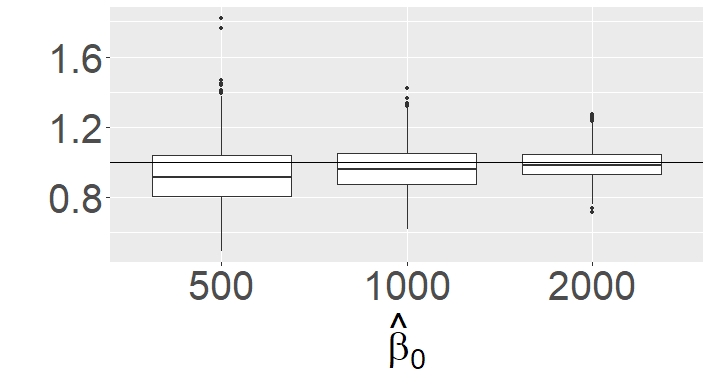}~\includegraphics[scale=0.3]{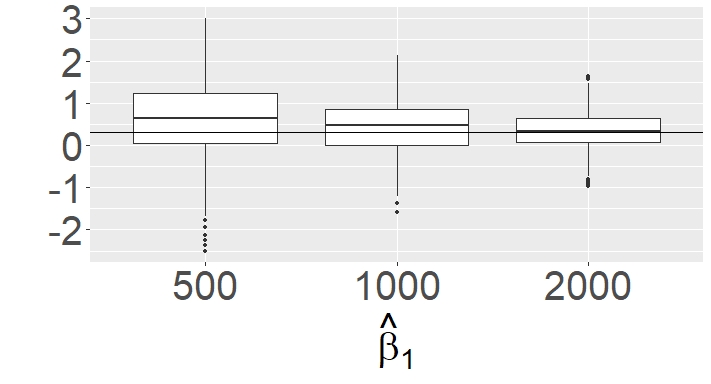}~\includegraphics[scale=0.3]{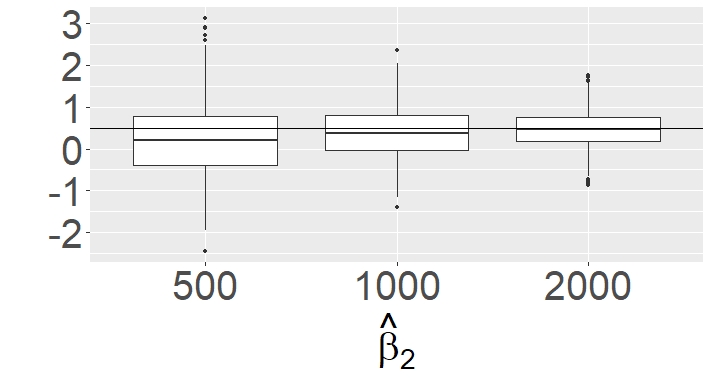}
		\includegraphics[scale=0.3]{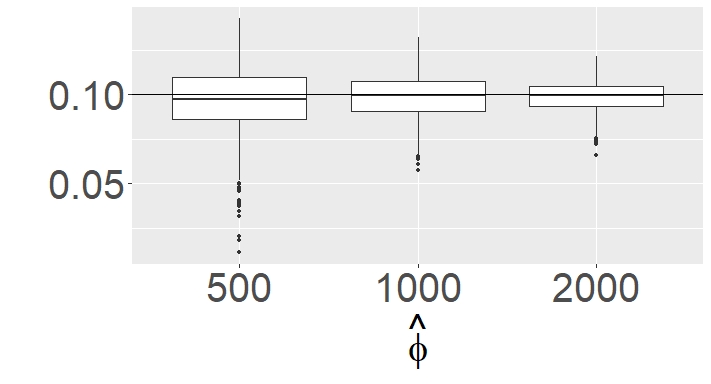}~\includegraphics[scale=0.3]{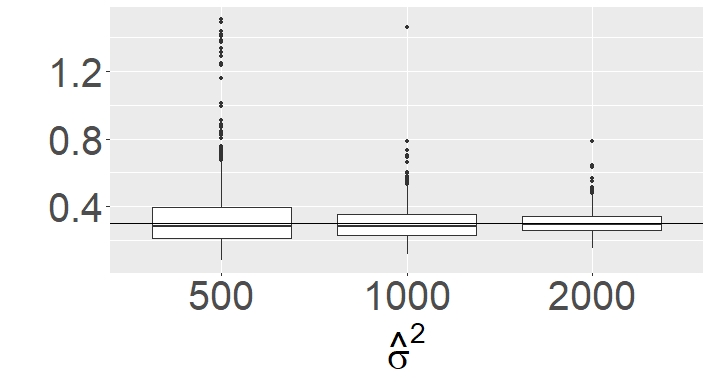}~\includegraphics[scale=0.3]{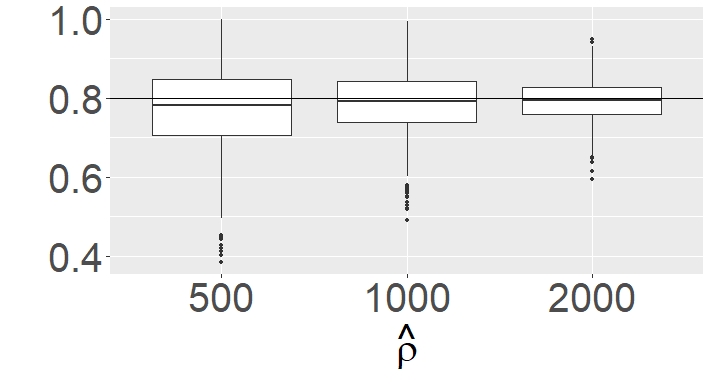}
		\caption{Boxplots of the parameter estimates based on the STS for bounded data.}
		\label{boxplot_bound}
	\end{figure}

	Figure \ref{boxplot_bound} shows the boxplots of the parameter estimates for the bounded time series case. From these plots, we have empirical evidence that the proposed estimators are consistent for the scenario considered here even for the quasi-likelihood estimators of the $\beta$'s.
	
	
	\begin{table}
		\centering
		\caption{Empirical means and standard errors of the quasi-likelihood estimates of $\beta$ and method of moments estimates of $\phi$, $\sigma^2$ and $\rho$ based on the STS for bounded data.}\label{R_bound} 
		\begin{tabular}{cccccccccc}
			\hline
			&&\multicolumn{2}{c}{$n=500$}&&
			\multicolumn{2}{c}{$n=1000$}&&
			\multicolumn{2}{c}{$n=2000$}\\
			\cline{3-4}\cline{6-7}\cline{9-10}
			parameter&true value&mean&stand. err.&& mean& stand. err.&& mean &stand. err.\\
			\hline
			$\beta_0$ &1   &0.932 &0.174 && 0.965&0.128 &&0.989 &0.090 \\
			$\beta_1$ &0.3 &0.616 &0.867 &&0.437 &0.608 &&0.349 &0.429 \\
			$\beta_2$ & 0.5&0.228 &0.869 &&0.384 &0.595 &&0.459 &0.423 \\
			$\phi$    & 0.1&0.096 &0.018 &&0.099 &0.012 &&0.099 &0.009 \\
			$\sigma^2$& 0.3&0.333 &0.201 &&0.301 &0.101 &&0.306 &0.069 \\
			$\rho$    & 0.8&0.773 &0.107 &&0.788 &0.079 &&0.792 &0.054 \\
			\hline
		\end{tabular}
	\end{table}

	\section{Time series data applications}\label{sec_app}
	
	In this section we apply the proposed STS models for analysing time series on unemployment rate and precipitation, so illustrating the performance of our bounded and positive continuous STS models in these data sets, respectively.
	
	\subsection{Unemployment rate data analysis}
	
\begin{figure}\centering
		\includegraphics[scale=0.7,height=5.5cm, width=7.5cm]{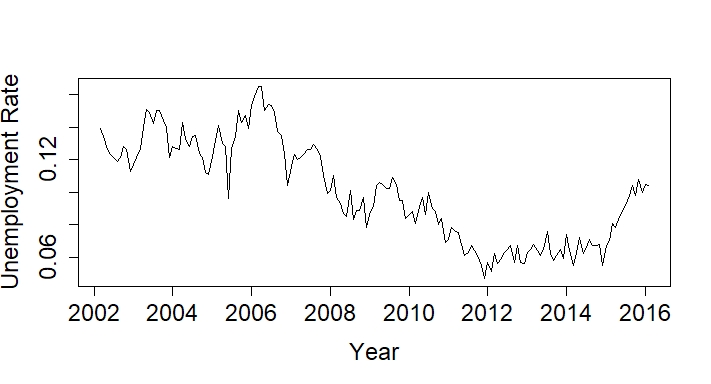}~\includegraphics[scale=0.7,height=5.5cm, width=7.5cm]{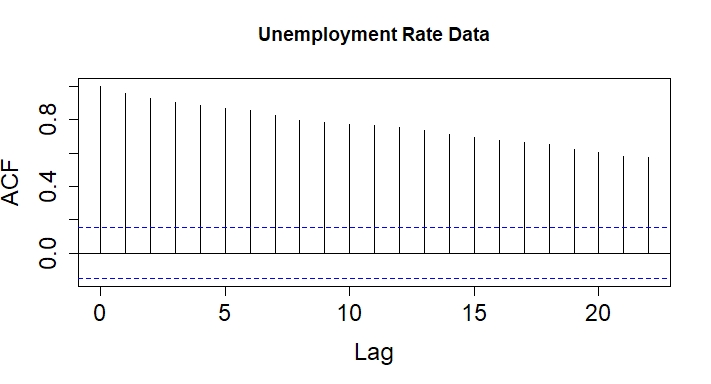}
		\caption{Plots of the monthly unemployment rate in the city of Recife from March 2002 to February 2016 (to the left) and its associated ACF (to the right).}
		\label{rate_app}
	\end{figure}

	This first application is devoted to the monthly unemployment rate in the city Recife/Brazil from March 2002 to February 2016 so totalizing $n=168$ observations, which can be obtained from website of the Institute of Applied Economic Research/Brazil (IPEA) \url{http://www.cbicdados.com.br/menu/emprego/pesquisa-mensal-de-emprego-ibge}. We here consider the bounded time series model presented in Subsection \ref{bound_case} for this application, where the data belongs to the unit interval $(0,1)$. Plots in Figure \ref{rate_app} display the unemployment rate time series and its associated ACF.
	
	From Figure \ref{rate_app}, we observe that this time series is non-stationary having a negative trend from March 2002 until time December 2011 ($t=118$). After this time, we observe a positive trend. To capture this behaviour, we consider the following covariate vector:
	\begin{equation*}
	x_{nt} = \left\{1,|t-118|/168 \right\}^\top, \quad t=1,\ldots,168.
	\end{equation*}
	
	Table \ref{fit_bound} provides the estimates of the parameters with their respective standard errors obtained through Monte Carlo simulation (fourth column). For the simulated results, we take a beta distribution for generating time series as discussed in Section \ref{sec_sim}. We also present the standard errors obtained from the quasi-likelihood estimation by ignoring the dependence among the observations due to latent process (second column). As it can be seen, there is a huge difference between the standard errors based on the Monte Carlo simulation (considering the presence of the latent process) and those ones from the quasi-likelihood approach. This is also nicely discussed on the papers by \cite{davis2000} and \cite{davis2009}, where a generalized linear model approach is considered.
	
	\begin{table}
		\centering
		\caption{Parameter estimates and respective standard errors of the semiparametric bounded time series model for the unemployment rate data.}\label{fit_bound} 
		\begin{tabular}{ccccccc}
			\hline
			&&\multicolumn{2}{c}{Quasi+MM}&&\multicolumn{2}{c}{Simulated}\\ 
			\cline{3-4} \cline{6-7} 
			covariates/par. && estimates        & stand.err.&& estimates        & stand.err. \\
			\hline
			Intercept       && 2.680            &0.027      &&2.896             & 0.121 \\
			$|t-118|/168$   && $-$1.213         &0.068      &&$-$2.208          & 0.242\\
			\hline
			$\phi$          && $1.4\cdot10^{-4}$&$-$        && $1.8\cdot10^{-4}$& $1.1\cdot10^{-4}$\\
			$\sigma^2$      &&      0.033       &$-$        &&0.053             & 0.024\\
			$\rho$          &&0.934             &$-$        &&0.893             & 0.031\\
			\hline
		\end{tabular}
	\end{table}

	From Table \ref{fit_bound}, it is also possible to note a good agreement between quasi-likelihood and method of moments estimates and those ones caught from Monte Carlo simulation, with exception of the trend coefficient. We have experienced this problem in our simulated results in the previous section. A possible solution for this will be discussed in the Concluding remarks Section. Anyway, this does not change inference about the associated covariate which is significant (sig. level at 5\%).
	
	\begin{figure}\centering
		\includegraphics[scale=0.4]{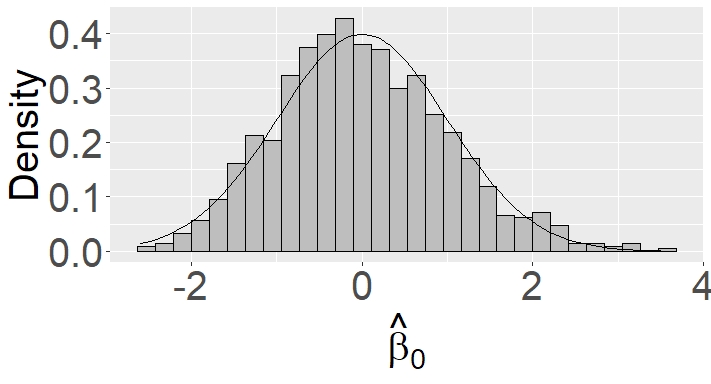}~\includegraphics[scale=0.4]{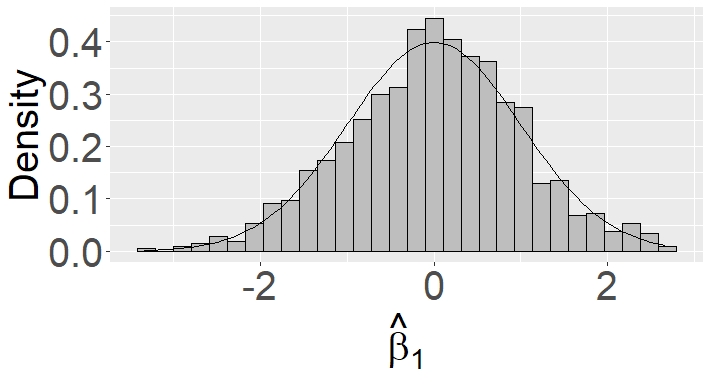}
		\caption{Histograms of the standardized quasi-likelihood estimate of the $\beta$'s for the unemployment data.}
		\label{fig_hist_unemp}
	\end{figure}
	
\begin{figure}\centering
		\includegraphics[scale=0.4]{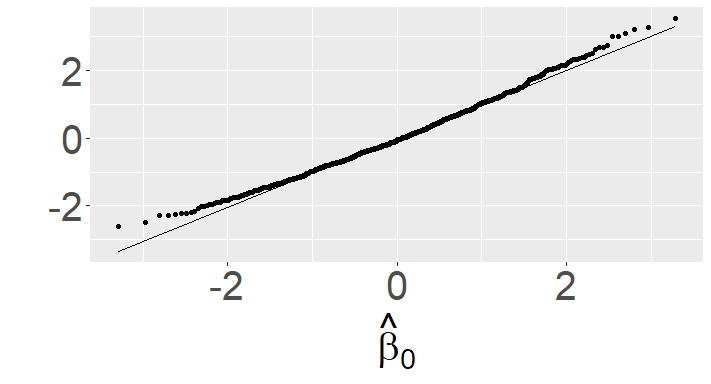}~\includegraphics[scale=0.4]{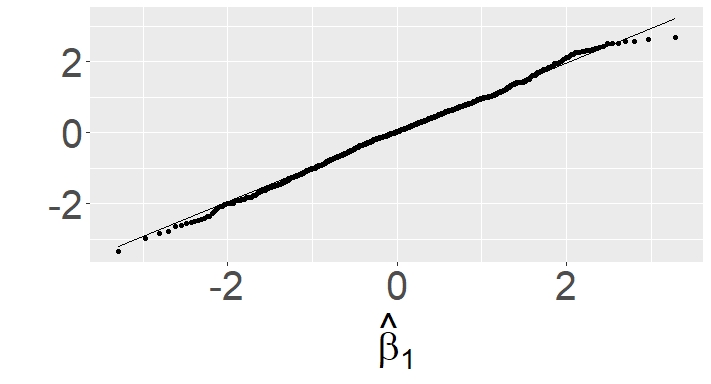}
		\caption{QQ-plots of the standardized quasi-likelihood estimate of the $\beta$'s for the unemployment data.}
		\label{fig_qq_unemp}
	\end{figure}

	Figures \ref{fig_hist_unemp} and \ref{fig_qq_unemp} respectively show the histograms and qq-plots of the standardized Monte Carlo estimates of the $\beta$'s, which indicate satisfactory normal approximations. 
	
	\subsection{Precipitation data analysis}
	
\begin{figure}\centering
		\includegraphics[scale=0.7,height=5.5cm, width=7.5cm]{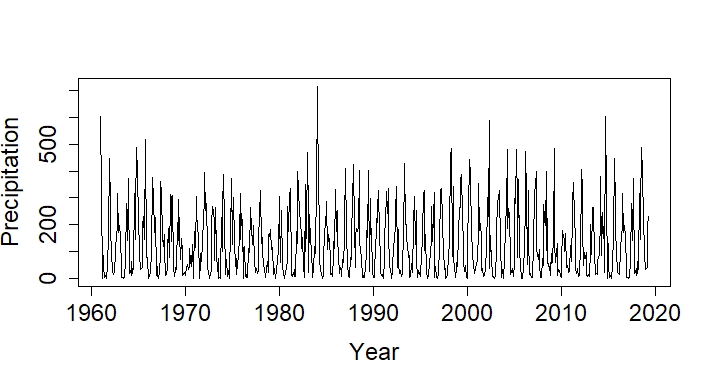}~\includegraphics[scale=0.7,height=5.5cm, width=7.5cm]{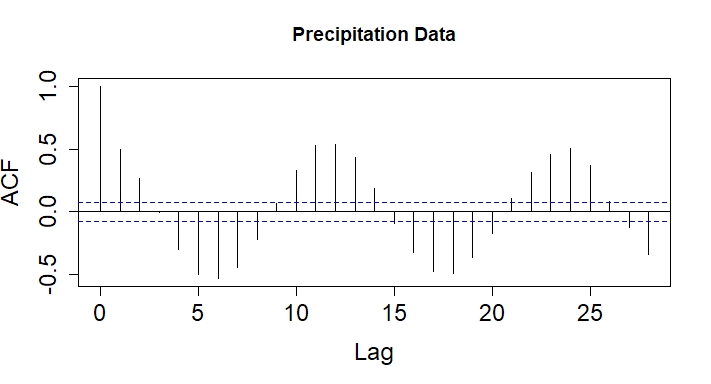}
		\caption{Plots of the monthly precipitation in the city of Juiz de Fora from January 1961 to May 2019 (to the left) and its associated ACF (to the right).}
		\label{app_prec}
	\end{figure}

	We now consider the monthly precipitation data (mm) of the city of Juiz de Fora in the state of Minas Gerais, Brazil, from January 1961 to May 2019. These data consist of $n=645$ observations and can be obtained from the Meteorological Database for Teaching and Research $-$ INMET, Brazil; please see \url{http://www.inmet.gov.br/portal/index.php?r=bdmep/bdmep}. Figure \ref{app_prec} presents the plots of the precipitation time series and its associated ACF. As expected, we can see a seasonal behaviour of this time series. The model for non-negative time series data given in Subsection \ref{non-neg_case} with $p=2$ ($V(\mu)=\mu^2$) is applied here. Following \citet{josrgensen2007}, we consider the following covariates for our analysis:
	\begin{equation*}
	\begin{aligned}
	\cos(2\pi t/j) & ,  \quad j = 12, 6, 4, 3, \\
	\sin(2\pi t/j) & ,  \quad j = 12, 6, 4, 3, 
	\end{aligned}
	\end{equation*}
	for $t=1,\ldots,645$.
	
	Table \ref{app_nonneg} shows the parameter estimates and respective standard errors of the STS model for the precipitation data. For getting the Monte Carlo results, we followed the strategy discussed in Section \ref{sec_sim} and considered a conditional gamma distribution for $Y_t$ given $\alpha_t$, for $t=1,\ldots,645$. 
	
	\begin{table}
		\centering
		\caption{Parameter estimates and respective standard errors of the semiparametric positive continuous time series model (with $p=2$) for the precipitation data.}\label{app_nonneg} 
		\begin{tabular}{ccccccc}
			\hline
			&&\multicolumn{2}{c}{Quasi-Likelihood}&&\multicolumn{2}{c}{Simulated}\\ 
			\cline{3-4} \cline{6-7} 
			covariates/par. && estimates        & stand.err.&& estimates        & stand.err. \\
			\hline
			Intercept&& 4.804   &0.038&&4.797   &0.063 \\
			$\cos(2\pi t/12)$&& $-$0.188&0.054&&$-$0.188&0.064\\
			$\sin(2\pi t/12)$&& 0.402   &0.054&&0.398   &0.065\\
			$\cos(2\pi t/6)$ && 0.065   &0.054&&0.063   &0.045\\
			$\sin(2\pi t/6)$ && 0.012   &0.054&&0.011   &0.045\\
			$\cos(2\pi t/4)$ && 0.040   &0.054&&0.042   &0.036\\
			$\sin(2\pi t/4)$ && $-$0.040&0.054&&$-$0.040& 0.036  \\
			$\cos(2\pi t/3)$ && $-$0.085&0.054&&$-$0.085&0.033\\
			$\sin(2\pi t/3)$ && 0.077   &0.054&&0.078   &0.032\\
			\hline
			$\phi$           && 0.031   &$-$  &&0.082   &0.060\\
			$\sigma^2$       && 0.525   &$-$  &&0.455   &0.093\\
			$\rho$           && 0.581   &$-$  &&0.626   &0.077\\
			\hline
		\end{tabular}
	\end{table}

	The quasi-likelihood and method of moments procedures provide similar estimates than the Monte Carlo method, specially for estimating the $\beta$'s. By using a significance level at 5\% and taking into account the latent process, the covariates $\cos(2\pi t/12)$, $\sin(2\pi t/12)$, $\cos(2\pi t/3)$ and $\sin(2\pi t/3)$ were significant. These covariates correspond to annual and quarterly seasonality. On the other hand, by ignoring the presence of the latent process, the quarterly seasonalities are not significant. This shows the importance of considering a suitable model specification, otherwise inference may be compromised.
	
	\begin{figure}
		\begin{subfigure}{.33\textwidth}
			\centering
			\includegraphics[width=\linewidth]{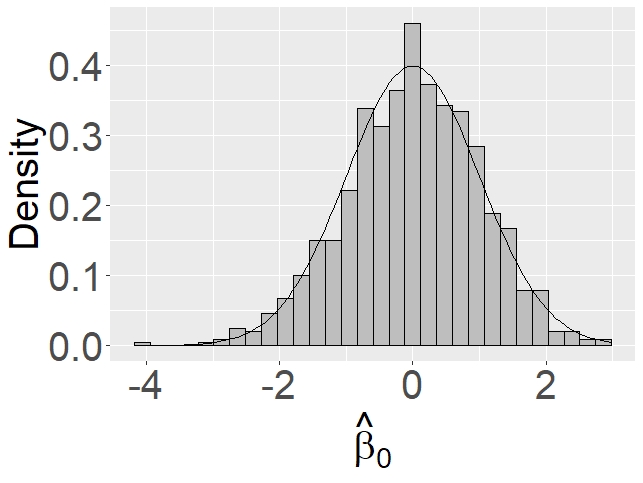}
		\end{subfigure}%
		\begin{subfigure}{.33\textwidth}
			\centering
			\includegraphics[width=\linewidth]{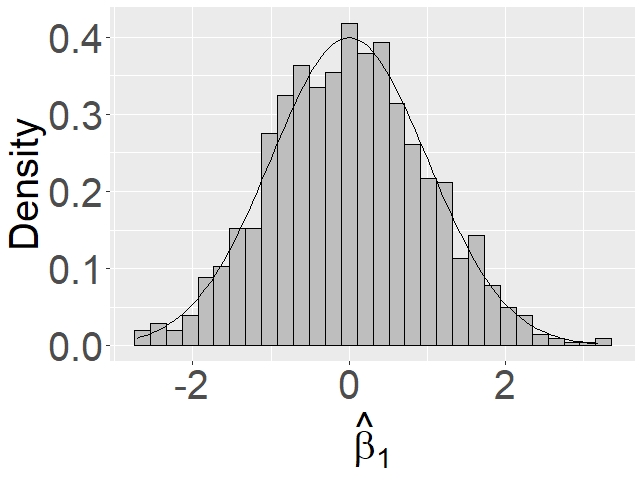}
		\end{subfigure}
		\begin{subfigure}{.33\textwidth}
			\centering
			\includegraphics[width=\linewidth]{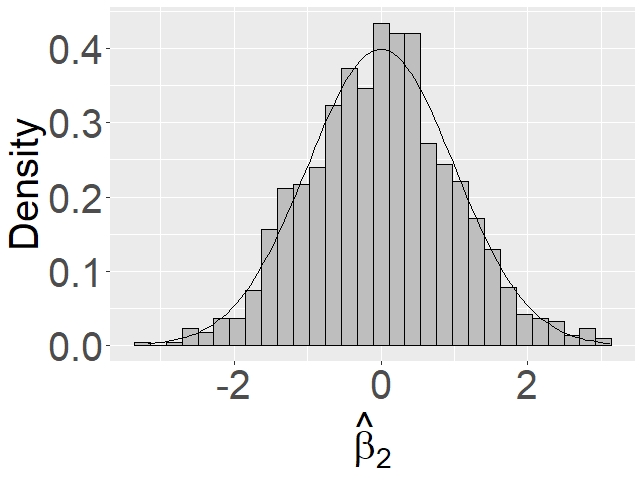}
		\end{subfigure}%
		
		\begin{subfigure}{.33\textwidth}
			\centering
			\includegraphics[width=\linewidth]{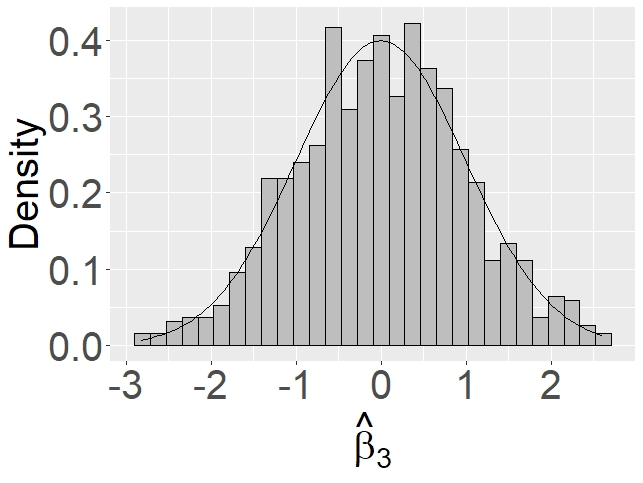}
		\end{subfigure}%
		\begin{subfigure}{.33\textwidth}
			\centering
			\includegraphics[width=\linewidth]{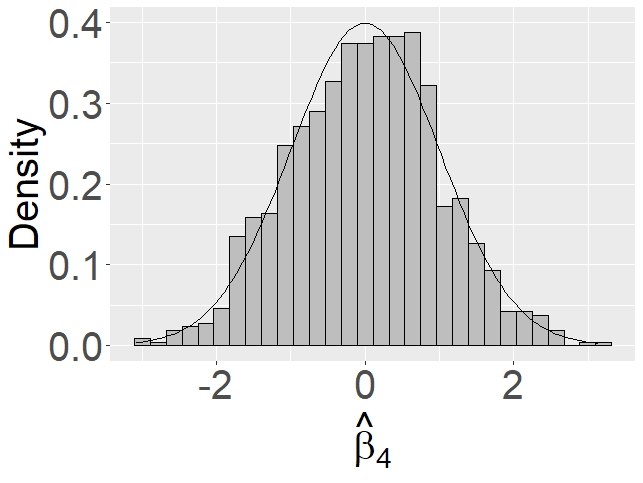}
		\end{subfigure}
		\begin{subfigure}{.33\textwidth}
			\centering
			\includegraphics[width=\linewidth]{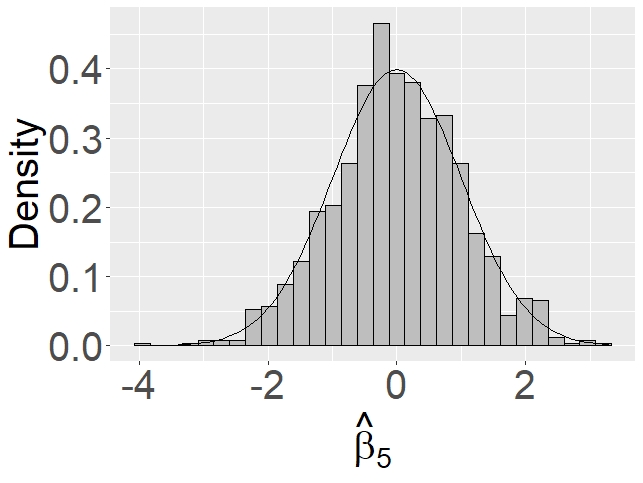}
		\end{subfigure}%
		
		\begin{subfigure}{.33\textwidth}
			\centering
			\includegraphics[width=\linewidth]{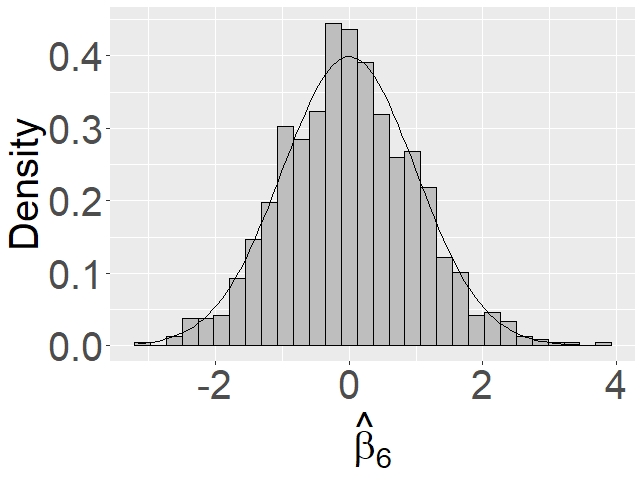}
		\end{subfigure}%
		\begin{subfigure}{.33\textwidth}
			\centering
			\includegraphics[width=\linewidth]{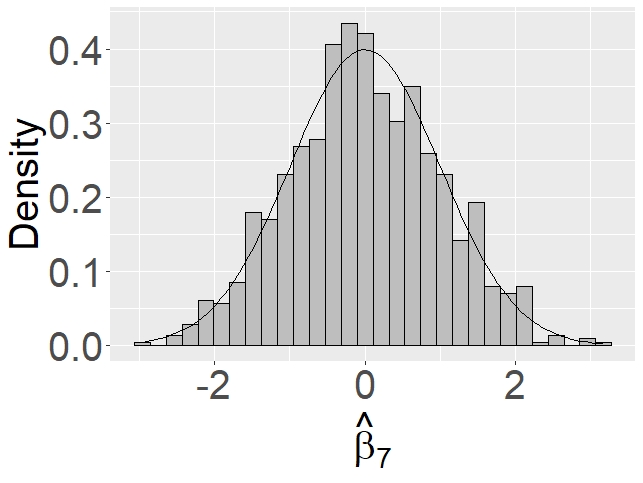}
		\end{subfigure}
		\begin{subfigure}{.33\textwidth}
			\centering
			\includegraphics[width=\linewidth]{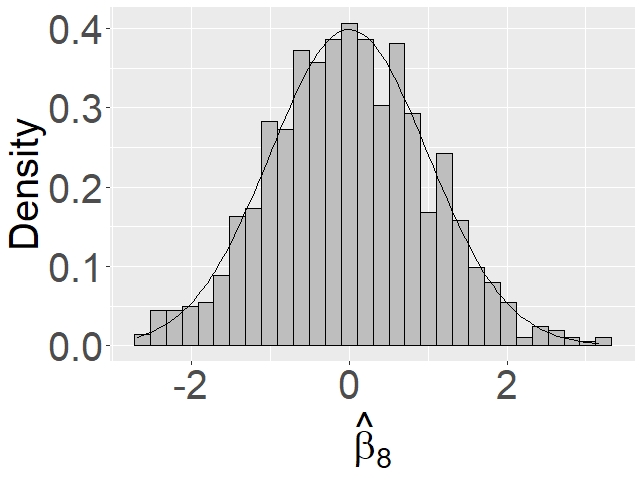}
		\end{subfigure}%
		\caption{Histograms of the standardized quasi-likelihood estimate of the $\beta$'s for the precipitation data.}
		\label{fig_hist_prec}
	\end{figure}

\begin{figure}
		\begin{subfigure}{.33\textwidth}
			\centering
			\includegraphics[width=\linewidth]{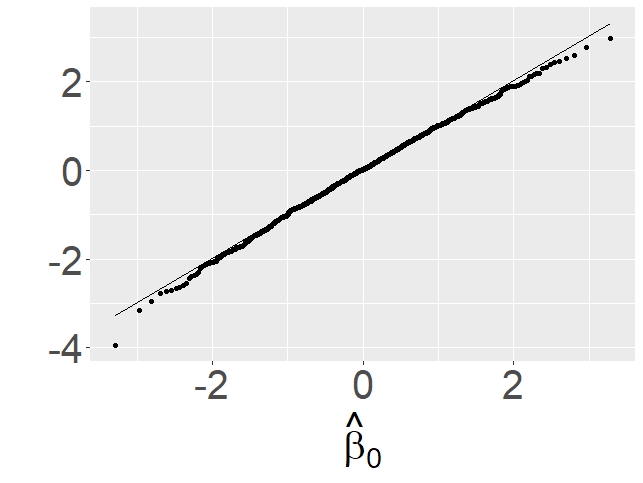}
		\end{subfigure}%
		\begin{subfigure}{.33\textwidth}
			\centering
			\includegraphics[width=\linewidth]{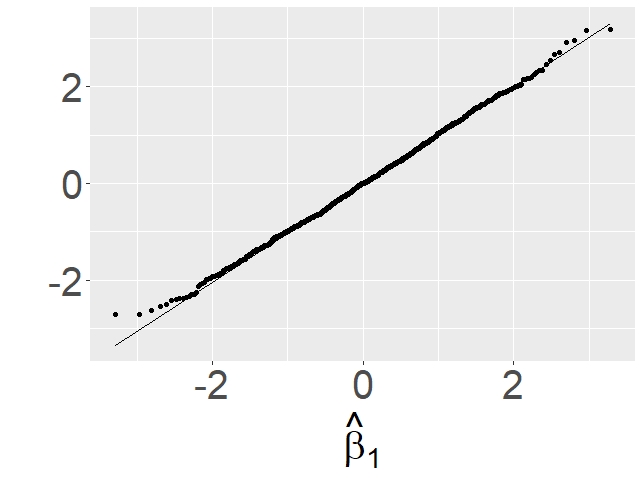}
		\end{subfigure}
		\begin{subfigure}{.33\textwidth}
			\centering
			\includegraphics[width=\linewidth]{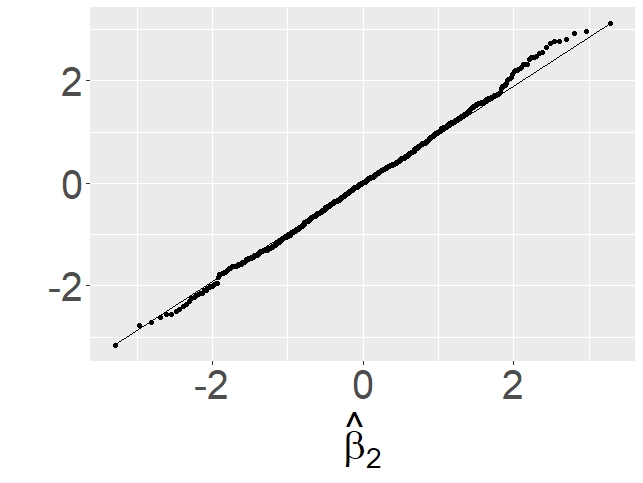}
		\end{subfigure}%
		
		\begin{subfigure}{.33\textwidth}
			\centering
			\includegraphics[width=\linewidth]{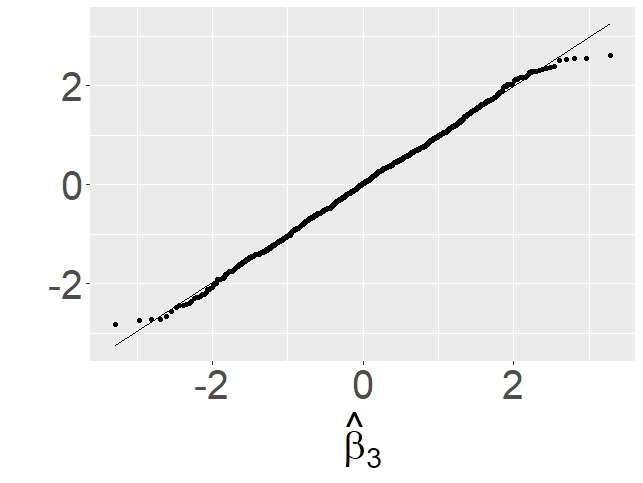}
		\end{subfigure}%
		\begin{subfigure}{.33\textwidth}
			\centering
			\includegraphics[width=\linewidth]{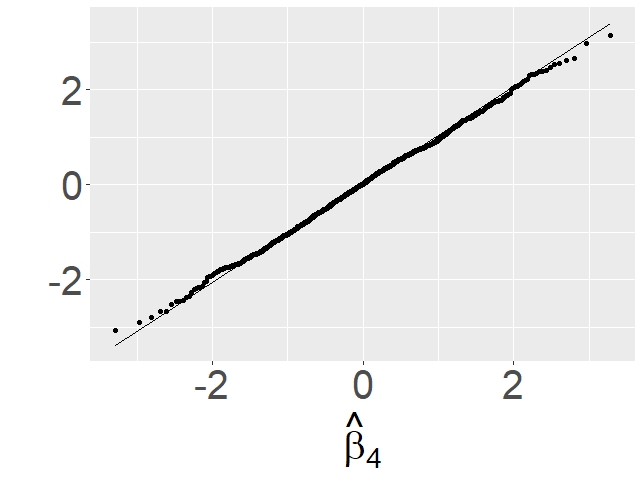}
		\end{subfigure}
		\begin{subfigure}{.33\textwidth}
			\centering
			\includegraphics[width=\linewidth]{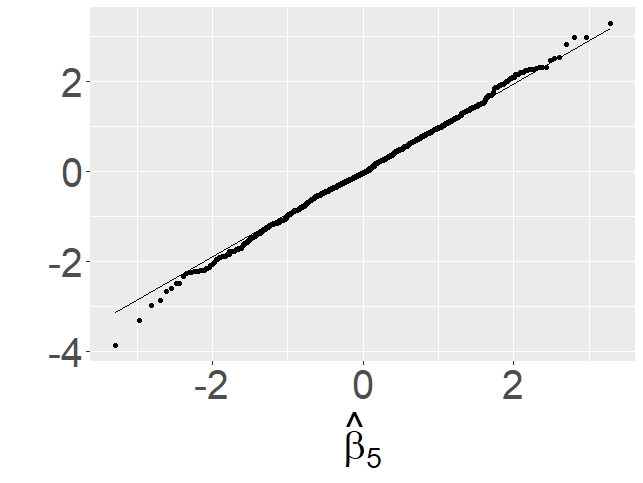}
		\end{subfigure}%
		
		\begin{subfigure}{.33\textwidth}
			\centering
			\includegraphics[width=\linewidth]{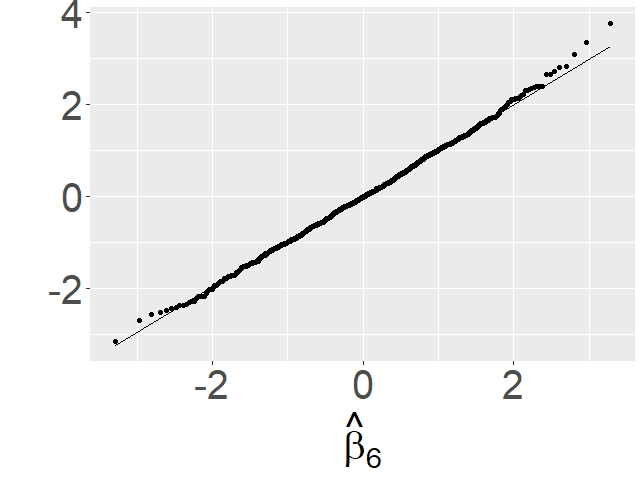}
		\end{subfigure}%
		\begin{subfigure}{.33\textwidth}
			\centering
			\includegraphics[width=\linewidth]{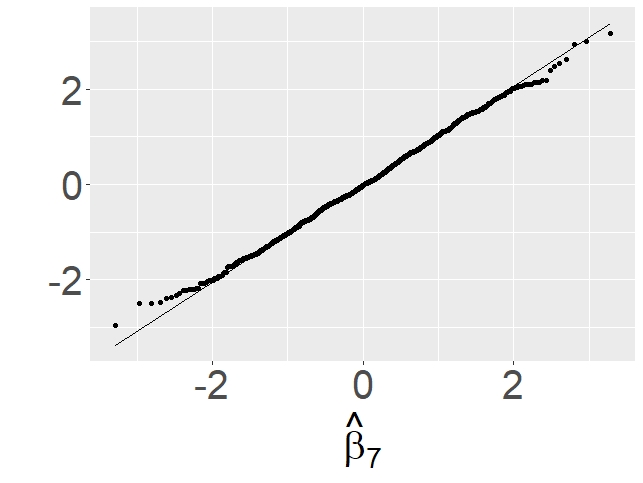}
		\end{subfigure}
		\begin{subfigure}{.33\textwidth}
			\centering
			\includegraphics[width=\linewidth]{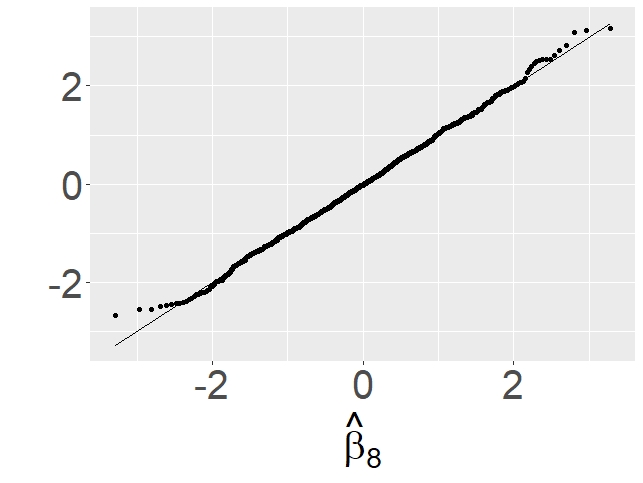}
		\end{subfigure}%
		\caption{QQ-plots of the standardized quasi-likelihood estimates of the $\beta$'s for the precipitation data.}
		\label{fig_qq_prec}
	\end{figure}

	In Figures \ref{fig_hist_prec} and \ref{fig_qq_prec}, we present the histograms and qq plots of the standardized quasi-likelihood estimates of the $\beta$'s, respectively. These plots again indicate a satisfactory normal approximation for the distribution of the quasi-likelihood estimators. This is in line with our simulated results provided in Section \ref{sec_sim}.
	
	\section{Concluding remarks}\label{concl_rem}
	
	A flexible class of semiparametric time series models was proposed by assuming a quasi-likelihood model driven by a latent factor process. Our proposed methodology is able for dealing with positive continuous, count, bounded, binary and real-valued time series. Inference on the model parameters was discussed and Monte Carlo simulations were addressed for checking estimation performance. Applications on unemployment rate and precipitation time series data illustrated the usefulness of the proposed methodology in practical situations.
	
	A challenging point seems to be the estimation of the parameters related to the mean for the bounded case, where a considerable bias was observed, which was also experienced by \cite{davis2009} in a binary time series model. A possible solution may be to use a Bootstrap procedure \citep{efrtib1994} for obtaining the bias and then correct the quasi-likelihood estimates. 
	
	Another point we would like to call attention is that other forms for the variance function can be considered and the results discussed in this paper can be easily adapted. For example, in the bounded case, one might be interested in considering the variance function $V(\mu)=\mu^3(1-\mu)^3$, with $\mu\in(0,1)$. The marginal moments and autocorrelation function for this case are obtained following the same steps given in Subsection \ref{bound_case}.
	
	Other points we believe that deserve to be investigated in future research are: (i) prediction; (ii) diagnostic tools and (iii) multivariate extension.
	
	\section*{Acknowledgments}
	\noindent G. Maia and W. Barreto-Souza would like to thank the financial support from Conselho Nacional de Desenvolvimento Cient\'ifico e Tecnol\'ogico (CNPq-Brazil, grant number 305543/2018-0). W. Barreto-Souza and H. Ombao would like to acknowledge support for their research by KAUST.
	
	\bibliographystyle{cas-model2-names}
	\bibliography{cas-refs}

	
	
	

\end{document}